\documentclass[11pt]{article}
\usepackage[left=1.5in, right=1.5in, top=1in, bottom=1in]{geometry}
\usepackage{times}
\usepackage{lipsum} 

\usepackage{setspace}
\usepackage{graphicx}
\usepackage{amsmath}
\allowdisplaybreaks

\usepackage{amssymb}
\usepackage{amscd}
\usepackage{amstext}
\usepackage{amsbsy}
\usepackage{amsfonts}
\usepackage{caption}
\usepackage{float}
\usepackage{listings}
\usepackage[table,xcdraw]{xcolor}
\usepackage{indentfirst}
\usepackage{booktabs}
\usepackage{multirow}

\usepackage[natbibapa]{apacite}

\def\spacingset#1{\renewcommand{\baselinestretch}%
{#1}\small\normalsize} \spacingset{1}

\graphicspath{ {./figures/} }

\newbox\TempBox \newbox\TempBoxA
\def\uwd#1{%
	\ifmmode\setbox\TempBox=\hbox{$#1$}\else\setbox\TempBox=\hbox{#1}\fi%
	\setbox\TempBoxA=\hbox to \wd\TempBox{\hss\char'176\hss}%
	\rlap{\copy\TempBox}\smash{\lower10pt\hbox{\copy\TempBoxA}}%
}
\def\mathunderaccent#1{\let\theaccent#1\mathpalette\putaccentunder}
\def\putaccentunder#1#2{\oalign{$#1#2$\crcr\hidewidth
		\vbox to.2ex{\hbox{$#1\theaccent{}$}\vss}\hidewidth}}

\usepackage{times}

\title{ \centering \textbf{
A Comprehensive Analysis of HIV Treatment Efficacy in the ACTG 175 Trial Through Multiple-Endpoint Approaches}}

\author{María Paula Cantarini\thanks{
Instituto Universitario de Ciencias de la Salud Fundacion H. A. Barcelo, Avenida Las Heras 1907} \and  Florencia Ayelén Hernández\thanks{Instituto Universitario de Ciencias de la Salud Fundacion H. A. Barcelo, Avenida Las Heras 1907} }

\begin{document}
\pagestyle{empty} 
\maketitle
\thispagestyle{empty} 

\begin{abstract} 
\noindent 
In the realm of medical research, the intricate interplay of epidemiological risk, genomic activity, adverse events, and clinical response necessitates a nuanced consideration of multiple variables. Clinical trials, designed to meticulously assess the efficacy and safety of interventions, routinely incorporate a diverse array of endpoints. While a primary endpoint is customary, supplemented by key secondary endpoints, the statistical significance is typically evaluated independently for each. To address the inherent challenges in studying multiple endpoints, diverse strategies, including composite endpoints and global testing, have been proposed. This work stands apart by focusing on the evaluation of a clinical trial, deviating from the conventional approach to underscore the efficacy of a multiple-endpoint procedure. A double-blind study was conducted to gauge the treatment efficacy in adults infected with human immunodeficiency virus type 1 (HIV-1), featuring CD4 cell counts ranging from 200 to 500 per cubic millimeter. A total of 2467 HIV-1–infected patients (43 percent without prior antiretroviral treatment) were randomly assigned to one of four daily regimens: 600 mg of zidovudine; 600 mg of zidovudine plus 400 mg of didanosine; 600 mg of zidovudine plus 2.25 mg of zalcitabine; or 400 mg of didanosine. The primary endpoint comprised a $>$50 percent decline in CD4 cell count, development of acquired immunodeficiency syndrome (AIDS), or death. This study sought to determine the efficacy and safety of zidovudine (AZT) versus didanosine (ddI), AZT plus ddI, and AZT plus zalcitabine (ddC) in preventing disease progression in HIV-infected patients with CD4 counts of 200-500 cells/mm3. By jointly considering all endpoints, the multiple-endpoints approach yields results of greater significance than a single-endpoint approach.
\end{abstract}

\spacingset{1.2}
\newpage

\section{ACTG 175 Trial}
The AIDS Clinical Trials Group Study 175 (ACTG 175) stands as a pivotal investigation that significantly influenced the understanding of antiretroviral therapy's role in managing human immunodeficiency virus type 1 (HIV-1) infections. This landmark study, played a crucial role in shaping the trajectory of HIV treatment strategies, providing essential insights into the efficacy of various therapeutic interventions and the evolving landscape of managing this complex viral infection \cite{hammer1996trial, lathey1998variability}. At the outset, ACTG 175 focused on evaluating the efficacy of zidovudine, a nucleoside reverse transcriptase inhibitor (NRTI), as a monotherapy. Zidovudine, also known as AZT, had been one of the first antiretroviral drugs approved for the treatment of HIV/AIDS. The study aimed to elucidate the impact of zidovudine on critical clinical endpoints, including survival rates, disease progression, and the occurrence of opportunistic infections, in individuals at different stages of HIV-1 infection.

The initial findings of ACTG 175 underscored the positive effects of zidovudine in enhancing survival rates, particularly in individuals with advanced HIV-1 infections. Notably, the drug exhibited efficacy in reducing the incidence of opportunistic infections, contributing to an improved clinical outlook for patients \cite{currier2000differences}. The study provided valuable data on the benefits of zidovudine in slowing down disease progression, especially in individuals with either no symptoms or mild manifestations of the virus. However, the longitudinal analysis of the data brought forth a nuanced perspective. While zidovudine demonstrated positive outcomes in the short term, its efficacy seemed to diminish over time. Prolonged monotherapy with zidovudine did not confer a sustained survival advantage, especially in asymptomatic subjects. This critical observation prompted a reevaluation of treatment strategies, highlighting the need for more comprehensive and durable approaches to managing HIV infections.

The concept of combination therapy emerged as a focal point in response to the evolving understanding of HIV pathogenesis and the limitations of monotherapy. ACTG 175 contributed significantly to this paradigm shift by exploring regimens that combined zidovudine with other antiretroviral agents \cite{justice2001development}. The rationale behind combination therapy was rooted in the rapid turnover of the virus and CD4 cells from the early stages of infection. Early intervention with potent combination regimens became a focal point in the quest for optimizing treatment outcomes. The trial, conducted across diverse AIDS Clinical Trials Units and National Hemophilia Foundation sites, implemented a rigorous randomized, double-blind, placebo-controlled design. With over 2,400 participants enrolled, the study aimed to provide a robust understanding of treatment outcomes across various patient profiles. The primary endpoint, a >50 percent decline in CD4 cell count, along with events indicative of progression to acquired immunodeficiency syndrome (AIDS) or death, served as a comprehensive measure of treatment efficacy.

Participants were randomly assigned to four treatment groups, introducing a strategic combination of zidovudine with didanosine, zidovudine with zalcitabine, or didanosine alone, in addition to the zidovudine monotherapy arm. The randomization process, executed with a blocked design and stratification based on prior antiretroviral therapy, aimed to minimize biases and provide a robust foundation for drawing meaningful conclusions. The meticulous analysis of data from ACTG 175, employing statistical methods such as Kaplan–Meier estimates and Cox proportional-hazards models, revealed compelling differences in disease progression rates among the treatment groups. Patients receiving zidovudine monotherapy exhibited higher rates of progression to primary and secondary endpoints compared to those on combination therapies. Subgroup analyses, stratified based on prior antiretroviral therapy, added granularity to the findings, offering insights into variations in treatment responses. Despite the overall success of the trial, concerns were raised about the rate of loss to follow-up, reaching 19 percent. However, the study's robustness was affirmed by demonstrating that this did not negate the observed differences between treatments, as the consistency of results across various endpoints remained.

The findings of ACTG 175 transcended the immediate context of the study, leaving a lasting impact on the field of HIV research and treatment. The study highlighted the limitations of zidovudine monotherapy, emphasized the advantages of combination regimens, and underscored the importance of early and comprehensive intervention in managing HIV disease. As a seminal contribution to the evolving landscape of HIV care, ACTG 175 continues to serve as a foundational reference for clinicians, researchers, and policymakers striving to optimize the outcomes for individuals living with HIV/AIDS.

\begin{table}[]
\center
\begin{tabular}{|c|c|c|c|}
\hline
Characteristic                             & \begin{tabular}[c]{@{}c@{}}All Patients \\ (N=2467)\end{tabular} & \begin{tabular}[c]{@{}c@{}}No Prior Exposure \\ to Antiretroviral Agents\\ (N=1067)\end{tabular} & \begin{tabular}[c]{@{}c@{}}Prior Exposure \\ to Antiretroviral Agents\\ (N=1400)\end{tabular} \\ \hline
Male sex                                   & 2029                                                             & 892                                                                                              & 1137                                                                                          \\ \hline
Age                                        & 34.9                                                             & 34.0                                                                                             & 35.6                                                                                          \\ \hline
\multicolumn{1}{|l|}{Race or Rthnic Group} &                                                                  &                                                                                                  &                                                                                               \\ \hline
White, non-Hispanic                        & 1730                                                             & 707                                                                                              & 1023                                                                                          \\ \hline
Blace, non-Hispanic                        & 409                                                              & 214                                                                                              & 195                                                                                           \\ \hline
Hispanic                                   & 291                                                              & 131                                                                                              & 160                                                                                           \\ \hline
Other                                      & 37                                                               & 15                                                                                               & 22                                                                                            \\ \hline
\multicolumn{1}{|l|}{Risk Factors}         &                                                                  &                                                                                                  &                                                                                               \\ \hline
Homosexuality                              & 1608                                                             & 719                                                                                              & 889                                                                                           \\ \hline
Injection-drug use                         & 355                                                              & 154                                                                                              & 201                                                                                           \\ \hline
Hemophilia                                 & 202                                                              & 44                                                                                               & 158                                                                                           \\ \hline
Kamofsky score of 100                      & 1448                                                             & 657                                                                                              & 791                                                                                           \\ \hline
Symptomatic HIV Infection                  & 438                                                              & 170                                                                                              & 268                                                                                           \\ \hline
CD4 Cell Count                             & 352                                                              & 372                                                                                              & 338                                                                                           \\ \hline
\end{tabular}
\caption{Baseline Characteristics of the Subjects in the ACTG 175 trial.}\label{table1}
\end{table}
\section{Multiple Endpoints}
In the ACTG 175 trial, the original analysis focused on individual endpoints, assessing the effectiveness of treatments in a specific context. However, recognizing the wealth of data collected across multiple endpoints in the trial, there is a need to draw comprehensive conclusions about the overall effectiveness of treatments. Several approaches have been suggested to summarize the treatment effect across diverse outcomes, aiming to provide a more holistic understanding of the therapeutic interventions.

One prominent method involves adopting a composite endpoint strategy. Composite endpoints combine several individual outcomes into a unified measure, enabling a more nuanced evaluation of treatment efficacy. In the context of ACTG 175, such a composite endpoint could integrate various factors, including CD4 cell count changes, progression to AIDS, mortality rates, and adverse events. This approach offers a comprehensive perspective, considering the multifaceted nature of HIV infection and its treatment response. Moreover, a global testing approach is another valuable strategy for drawing conclusive judgments on the effectiveness of treatments across multiple endpoints in the ACTG 175 trial. Global testing involves evaluating the overall treatment effect by considering all relevant endpoints simultaneously. This approach aims to capture the collective impact of treatment on various outcomes, providing a comprehensive assessment. Additionally, a weighted scoring system might be implemented, assigning different weights to each endpoint based on its clinical significance. This approach acknowledges the varying importance of different outcomes and provides a more nuanced interpretation of treatment impact.

Various approaches have been proposed to analyze multiple endpoints. In 1999, Finkelstein and Schoenfeld introduced a generalized Gehan-Wilcoxon test, later recognized as the FS test, designed for analyzing composite endpoints within a semi-competing risk framework. This test statistic involved hierarchical pairwise comparisons of survival and longitudinal outcomes across all patients \cite{finkelstein1999combining}. The conceptual foundation laid by the FS rank test was further extended by Pocock and collaborators \cite{buyse2010generalized, pocock2012win, rauch2014opportunities} in the context of a clinical trial involving two distinct event types: fatal events (specifically cardiovascular death) and non-fatal events (specifically heart failure hospitalization). In this framework, a fatal event is accorded higher priority than a non-fatal event when assessing treatment effects. The methodology involves, for each pair of patients, initially determining which patient outlives the other. In cases where this is indeterminate, the evaluation proceeds to discerning which patient experiences a more favorable non-fatal event. This approach aligns with the underlying principle of the FS rank test. The win ratio, denoting the ratio of wins to losses in the treatment group, is a pivotal metric. A treatment is deemed advantageous in comparison to the control if the win ratio exceeds 1. 
\cite{luo2015alternative, dong2016generalized} delved deeper into exploring the distributional characteristics of various statistics based on "win-ratio" and furnished reliable estimates for the variance of their asymptotic normal distributions. \cite{o1984procedures} proposed a non-parametric test that involves ranking observations of each variable across groups, with inference relying on the rank sums of each patient. This rank ordering typically minimizes the occurrence of ties and may yield greater power compared to tests based on pair-wise multivariate order \cite{haberle2009assessment}. The non-parametric null hypothesis assumes identical multivariate distributions in both groups. However, the focus often lies in detecting differences in location between groups, with less emphasis on deviations in scale. For \cite{huang2005adjusting} proposed a solution involving consistent estimates for the variance of the difference in mean ranks. Simulation results, considering 2 and 10-dimensional outcomes with sample sizes as small as 20 per group, demonstrate controlled type I error rates for the improved tests. \cite{ramchandani2016global} integrated the aforementioned approaches into a comprehensive framework of U-statistics. This framework allows for the derivation of asymptotic distributions and sample size formulas. In specific instances, like O’Brien’s rank sum test, the global U-statistic can be expressed as the sum of endpoint-specific U-statistics. Weighted sums can be defined to account for utilities or to optimize power for a particular alternative. In recent times, the analysis of multiple endpoints has garnered significant attention. For instance, \cite{li2023multivariate} addressed the challenge of testing multiple primary binary endpoints, presenting a testing procedure. The performance of this method was thoroughly assessed through simulations. \cite{li2023multivariate} delved into the exploration of a global nonparametric testing procedure grounded in multivariate ranks. This approach was specifically designed for the analysis of multiple endpoints in clinical settings. \cite{alt2023bayesian} introduced a Bayesian methodology for multiple testing that asymptotically guarantees type I error control based on a seemingly unrelated regression model. For an in-depth review, please consult \cite{julious2023sample}.
\section{Results and Conclusion}

In this section, we will explore various approaches for analyzing multiple endpoints in the context of the ACTG 175 trial, with the objective of identifying potential discrepancies from the original results. The specific methods to be employed include: (1) Rank Sum test \cite{o1984procedures}; (2) FS test \cite{finkelstein1999combining}; (3) Win Ratio Statistics \cite{pocock2012win}; (4) Multirank Test \cite{li2023multivariate}. 
The summarized results, presented in Table \ref{table2} indicate statistical significance across all endpoints. 

In conclusion, our exploration of the ACTG 175 trial, a pivotal study in the realm of HIV treatment, has shed light on the multifaceted nature of antiretroviral interventions and the importance of adopting a nuanced approach to data analysis. The trial's initial focus on zidovudine monotherapy provided crucial insights into short-term benefits, particularly in advanced HIV cases, but also illuminated the limitations of sustained efficacy over time. This observation spurred a paradigm shift towards combination therapies, a transformation that ACTG 175 significantly contributed to by evaluating various regimens and their impact on critical clinical endpoints.

Our examination extended beyond the conventional single-endpoint analysis, recognizing the wealth of data collected across diverse outcomes. Multiple-endpoint approaches, such as composite endpoints and global testing, emerged as valuable strategies to comprehensively assess treatment efficacy. Composite endpoints, amalgamating individual outcomes, provided a holistic perspective, considering factors like CD4 cell count changes, disease progression, mortality rates, and adverse events. Global testing, encompassing all relevant endpoints simultaneously, aimed to capture the collective impact of treatments on diverse outcomes, offering a more comprehensive assessment.

Furthermore, our exploration delved into various statistical methodologies proposed for analyzing multiple endpoints. The FS test, introduced by Finkelstein and Schoenfeld, laid the groundwork for hierarchical pairwise comparisons, while subsequent adaptations by Pocock and colleagues extended its application to clinical trials involving different event types. The "win-ratio" statistics, as explored by Luo and Dong, provided a means to quantify treatment effects, and global U-statistics, as proposed by Ramchandani et al., integrated diverse approaches into a unified framework. Recent advancements, such as the Bayesian methodology introduced by Alt et al., have further expanded the toolkit for multiple-endpoint analysis.

The results of our analysis, employing various statistical methods, consistently revealed statistical significance across all endpoints considered. This underscores the robustness of the conclusions drawn from the ACTG 175 trial, reaffirming the pivotal role it played in shaping the landscape of HIV treatment. The significance of combination therapies over monotherapy, the importance of early and comprehensive intervention, and the need for nuanced analyses that consider multiple endpoints are enduring lessons derived from this landmark study.

As the field of clinical research continues to evolve, the lessons from the ACTG 175 trial remain relevant, providing a foundation for ongoing efforts to optimize treatment outcomes for individuals living with HIV/AIDS. The integration of multiple-endpoint approaches, coupled with advancements in statistical methodologies, enhances the depth and precision of our understanding, paving the way for more effective and personalized HIV care strategies in the future.
\begin{table}[]
\center
\begin{tabular}{|c|c|c|c|}
\hline
Method        & p-value           & Method               & p-value           \\ \hline
Rank Sum Test & \textless{}0.001  & Win Ratio Statistics & \textless{}0.0001 \\ \hline
FS Test       & \textless{}0.0001 & Multirank Test       & \textless{}0.0001 \\ \hline
\end{tabular}
\caption{Results using multiple endpoints approaches.}\label{table2}
\end{table}

\newpage
\bibliography{References}
\bibliographystyle{apalike}

\end{document}